\documentclass[aps,prc,10pt,twocolumn,amsmath,amssymb,superscriptaddress]{revtex4-2}

\usepackage{graphicx}
\usepackage{color}
\usepackage{float}
\newcommand{\be}{\begin{equation}}
\newcommand{\ee}{\end{equation}}
\newcommand{\beq}{\begin{eqnarray}}
\newcommand{\eeq}{\end{eqnarray}}
\newcommand{\ba}{\begin{array}}
\newcommand{\ea}{\end{array}}

\begin{document}

\title{Search for $\Theta^+$ in $K_L p \to K^+ n$ reaction in KLF at JLab}
\vspace{3mm}
\date{\today}

%--------------------- AUTHOR LIST ---------------------------
%\input{author_list.tex}
\author{\mbox{Moskov~J.~Amaryan}}
\altaffiliation{\texttt{mamaryan@odu.edu}}
\affiliation{Department of Physics, Old Dominion University, Norfolk, Virginia 23529, USA}

\author{\mbox{Shu~Hirama}}
\altaffiliation{\texttt{hirama.s.aa@m.titech.ac.jp}}
\affiliation{Department of Physics, Tokyo Institute of Technology, Megro, Tokyo 152-8551, Japan}

\author{\mbox{Daisuke~Jido}}
\altaffiliation{\texttt{jido@th.phys.titech.ac.jp}}
\affiliation{Department of Physics, Tokyo Institute of Technology, Megro, Tokyo 152-8551, Japan}

\author{\mbox{Igor~I.~Strakovsky}}
\altaffiliation{\texttt{igor@gwu.edu}}
\affiliation{Institute for Nuclear Studies, Department of Physics, 
    The George Washington University, Washington, DC 20052, USA}

%---------------------- ABSTRACT -----------------------------
\begin{abstract}
The possibility of the existence of multiquark hadrons made of 4-quark for mesons and 5-quark for baryons was predicted by Gell-Mann in Ref.~\cite{Gell-Mann:1964ewy}. The renewed interest for the search of exotic pentaquark states was initiated by the paper by Diakonov, Petrov, and Polyakov in Ref.~\cite{Diakonov:1997mm}. The 2003 experimental reports on the observation of $\Theta^+$ pentaquark with a $uudd\bar{s}$ quark content created a big excitement and many following experiments have reported its observation~\cite{Hicks:2012zz}.

After high-statistics experiments at JLab, which did not confirm previous claims by the CLAS collaboration, the community concluded that the $\Theta^+$ pentaquark either does not exist at all or has an extremely small cross section, making it currently unobserved.

There were different review papers on this subject, either questioning the existence of the $\Theta^+$ or attempting to explain the reasons why reaching a conclusion based on production experiments is 
challenging~\cite{Amaryan:2022iij}.

To address the challenge of minimal 3-body final states, a formation experiment with a projectile kaon beam is proposed. Below, we discuss how the $\Theta^+$ could be observed in the $K_L p \to \Theta^+ \to K^+ n$ reaction in the KLF experiment at JLab~\cite{KLF:2020gai}.
\end{abstract}

\maketitle

%------------------------------------------------------------
%\clearpage
\section{Introduction}

QCD gives rise to the hadron spectrum~\cite{Gell-Mann:1964ewy} and many $q\bar{q}$ and $qqq$ have been observed~\cite{ParticleDataGroup:2022pth}. However, the $q\bar{q}q\bar{q}$ and $qqqq\bar{q}$ and other many  quark  states are not forbidden either. Recently, LHCb collaboration claims evidence for four hidden-charm $qqqq\bar{q}$  
states near open-charm decay thresholds for $\Sigma_c^+\bar{D}^0$ and $\Sigma_c^+\bar{D}^{\ast 0}$ in $\Lambda_b \to J/\psi p K^-$ decays~\cite{LHCb:2019kea}. Nevertheless, although there is no doubt about LHCb observations, these states are not manifestly multi-quark states and there is a room that their flavor can be explained by three quarks. In the light quark sector, there is a clearly exotic $\Theta^+ (uudd\bar{s})$ state, yet to be unequivocally observed and identified. This particle along with other members of $\bar {10}$ has been proposed in Ref.~\cite{Diakonov:1997mm} with a mass of $M_{\Theta^+}=1.53~\mathrm{GeV}$ and a width less than $15~\mathrm{MeV}$. Due to relatively low mass and simple decay channels to $K^+n$ or $K^0p$ it has attracted attention of many experimental collaborations at different facilities worldwide (see, for instance, review papers~\cite{Hicks:2012zz, Amaryan:2022iij}). 
The initial experimental evidence for $\Theta^+$ was presented by the LEPS Collaboration at SPring-8~\cite{LEPS:2003wug} and the DIANA Collaboration from ITEP~\cite{DIANA:2003uet}. Subsequently, several experimental groups declared the observation of $\Theta^+$ but later retracted their claims. However, there are still persistent claims from other groups, maintaining the existence of $\Theta^+$ in an inconclusive status. Luckily there is a  unique possibility to use intensive incoming kaon beams to observe it in a formation experiments in a 2-body reactions free from reflections of other states simultaneously produced in a many-body final states.

%------------------------------------------------------------
%\clearpage
\section{KLF Experiment in Hall~D at JLab}
\label{Sec:Exp}

The K-Long project, led by the K-Long Facility (KLF) Collaboration, has been approved for a 200-day run. This project necessitates the establishment of a secondary  $K_L$ beamline in Hall~D at Jefferson Lab. Boasting a flux of $10^4~\mathrm{K_L/s}$ on a physics target—three orders of magnitude higher than previously achieved at SLAC (refer to Ref.~\cite{KLF:2020gai} and citations therein) — the KLF experiment stands out as a unique facility. The KLF has been specifically mentioned in the Long Range Plan of Department of Energy Office of Science (LPR2023) as the future physics program in Hall~D with the GlueX 
setup~\cite{LRP:2023}.

By conducting scattering experiments on both proton and neutron targets, the K-Long Facility will distinguish itself as the world's first secondary neutral kaon beam facility with a sufficiently high beam intensity to elucidate all $\Sigma^\ast$ and $\Lambda^\ast$ resonances up to $M = 2500~\mathrm{MeV}$ in formation reactions using precise partial wave analysis (PWA) to determine the pole positions and the widths of these resonances. While proton accelerators produce neutral kaons with significantly higher intensity, the simultaneous generation of orders of magnitude more neutrons restricts their utility to the search for rare decays, as seen in KOTO experiment at J-PARC~\cite{koto:2006}, rather than for spectroscopy. This capability is achievable only with the high-intensity $I = 5~\mathrm{\mu A}$ electron beam of CEBAF at JLab, uniquely being able to deliver a beam with a $64~\mathrm{ns}$ (or even $128~\mathrm{ns}$) bunch separation by running it  in a storage ring mode.

The KLF experiment aims to uncover ``missing'' hyperons~\cite{Koniuk:1979vw} through reactions, exemplified in Table~\ref{tbl:res1}. Additionally, it will conduct strange meson spectroscopy to determine pole positions in the I = 1/2 and 3/2 channels.
%---------------------------------------------------------
\begin{table}[htb!]
\vspace{-0.3cm}
\centering \protect\caption{The list of reactions which allow to study some of hyperon 
resonances~\cite{KLF:2020gai}.}

\vspace{2mm}
{%
\begin{tabular}{|c|c|}
\hline
Hyperon        & Reaction   \\
\hline
$\Sigma^\ast$  & $K_Lp\to \pi\Sigma^\ast \to \pi\pi\Lambda$ \\
$\Lambda^\ast$ & $K_Lp\to \pi\Lambda^\ast \to \pi\pi\Sigma$ \\
$\Xi^\ast$     & $K_Lp\to K\Xi^\ast \to \pi K\Xi^\ast$ \\
$\Omega^\ast$  & $K_Lp\to K^+K^+\Omega^\ast$ \\
\hline
\end{tabular}} \label{tbl:res1}
\end{table}
%--------------------------------------------
\begin{figure*}[htb!]
\vspace{-0.3cm}
\centering
{
    \includegraphics[width=0.75\textwidth,keepaspectratio]{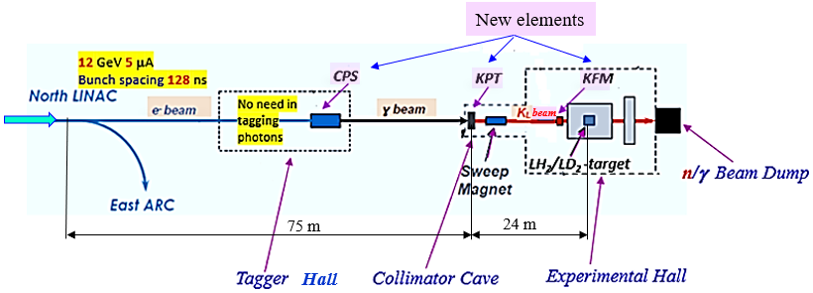} 
}

\centerline{\parbox{0.7\textwidth}{
\caption[] {\protect\small
Schematic view of the KLF beam line in JLab Hall~D with the production chain $e\rightarrow\gamma\rightarrow K_L$. The main components are the CPS, KPT, sweep magnet, and KFM. The beam goes from the left to the right~\cite{KLF:2020gai}.} 
\label{fig:beam1} } }
\end{figure*}
%--------------------------------------------

Besides the ordinary hyperons composed of 3-quarks, the KLF experiment may also observe the exotic $\Theta^+$ pentaquark, located at the apex of the anti-decuplet, $\bar {10}$, triangle. Furthermore, the experiment is sensitive to the observation of the exotic $\Xi^+$, positioned in the lower right corner of the $\bar {10}$ triangle, in the reaction $K_Lp \to K_S\Xi^+$ as it was noticed in ~\cite{Thiel:2020eqn}. It is also designed to detect another exotic state, $\Xi^{--}$, situated in the lower left corner of the $\bar {10}$ triangle, within the reaction $K_Lp\to K^+K^+K^+\Xi^{--}$. The 4~$\pi$ acceptance of the GlueX setup is well suited to serve a purpose.

Hall~D is acquiring a new basic equipment for the KLF project~\cite{KLF:2020gai}: the Compact Photon Source (CPS) situated in the Tagger Hall, the Kaon Production Target (KPT) located in the Collimator Cave, and the Kaon Flux Monitor (KFM) positioned in the Experimental Hall just in front of the GlueX spectrometer. A schematic view of the setup, including the GlueX spectrometer, is presented in Fig.~\ref{fig:beam1}. According to the current schedule, the KLF experiment will build all beamline components in 2024, install them in 2025, and commence data collection in 2026.

\begin{figure}[htb!]
\vspace{-0.3cm}
\centering
{
    \includegraphics[width=0.45\textwidth,keepaspectratio]{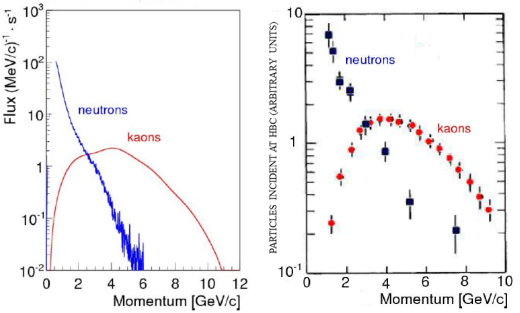} 
}

\centerline{\parbox{0.4\textwidth}{
\caption[] {\protect\small
\underline{Left}: Rate of $K_L$ (red) and neutrons (blue) on the LH$_2$/LD$_2$ cryogenic target of Hall~D at $12~\mathrm{GeV/c}$ as a function of their generated momenta, with a total rate of $1 \times 10^4~\mathrm{K_L/sec}$ and $6.6 \times 10^5~\mathrm{n/sec}$, respectively~\cite{KLF:2020gai}. 
\underline{Right}: Experimental data from SLAC measurements using a $16~\mathrm{GeV}$ electron 
beam~\cite{Brandenburg:1972pm}.
} 
\label{fig:yield1} } }
\end{figure}

Figure~\ref{fig:yield1} demonstrates that KLF simulations for the KLF kaon and neutron flux at $12~\mathrm{GeV}$ (left) are in a reasonable agreement with the $K_L$ and neutron spectra measured by SLAC at $16~\mathrm{GeV}$~\cite{Brandenburg:1972pm} (right).

%--------------------------------------------
%--------------------------------------------

\begin{figure}[htb!]
%\vspace{-0.5cm}
\centering
{
    \includegraphics[width=0.3\textwidth,keepaspectratio]{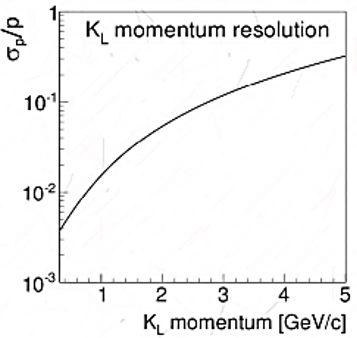} 
}

\centerline{\parbox{0.4\textwidth}{
\caption[] {\protect\small
 Momentum resolution ($\sigma_p/p$) as a function of the initial kaon momentum~\cite{KLF:2020gai}.} 
\label{fig:sigma1} } }
\end{figure}

The momentum of a $K_L$ beam will be measured using time-of-flight (TOF), specifically the time between the accelerator bunch (RF signal from CEBAF) and the reaction in the cryogenic LH$_2$ (LD$_2$) target, as detected by the GlueX spectrometer. The TOF resolution is the quadratic sum of the accelerator time and GlueX spectrometer time resolutions. Given the excellent time resolution of the accelerator signal, on the order of a few picoseconds, the TOF resolution will be predominantly determined by the GlueX detector. In our calculations, we utilized the currently achieved Start Counter time resolution of $250~\mathrm{psec}$ to illustrate the beam momentum resolution versus kaon momentum (Fig.~\ref{fig:sigma1}).

%--------------------------------------------

%------------------------------------------------------------
\vspace{-0.5cm}
\section{Expected statistics for the reaction $K_L p \to  \Theta^+ \to K^+ n$}
\label{Sec:stat}

Utilizing a modified Partial Wave Analysis (PWA), Arndt and co-workers conducted a re-analysis of the existing $KN$ database~\cite{Hyslop:1992cs}. The aim was to investigate the impact of a narrow state on fits to $K^+-N$ observables~\cite{Arndt:2003xz}. The study revealed that the presence of a $\Theta^+$ in the $P_{01}$ ($J^P = 1/2^+$) state, the mass of $\sim1545~\mathrm{MeV}$ and with a width of $\Gamma(\Theta^+) \le 0.5~\mathrm{MeV}$, is conceivable (refer to Fig.1(b) in Ref.\cite{Arndt:2003xz}).

The charge-exchange reaction $K^+Xe\to K^0pXe'$ was investigated using the data of the DIANA 
experiment~\cite{DIANA:2013mhv}. Using the ratio between the numbers of resonant and non-resonant charge-exchange events in the peak region, the intrinsic width of this baryon resonance is determined
as $\Gamma(\Theta^+) = 0.34 \pm 0.10~\mathrm{MeV}$.

In our study, we will consider a width of $\Gamma(\Theta^+) = 0.4~\mathrm{MeV}$.

\begin{figure}[htb!]
%\vspace{-0.5cm}
\centering
{
    \includegraphics[width=0.45\textwidth,keepaspectratio]{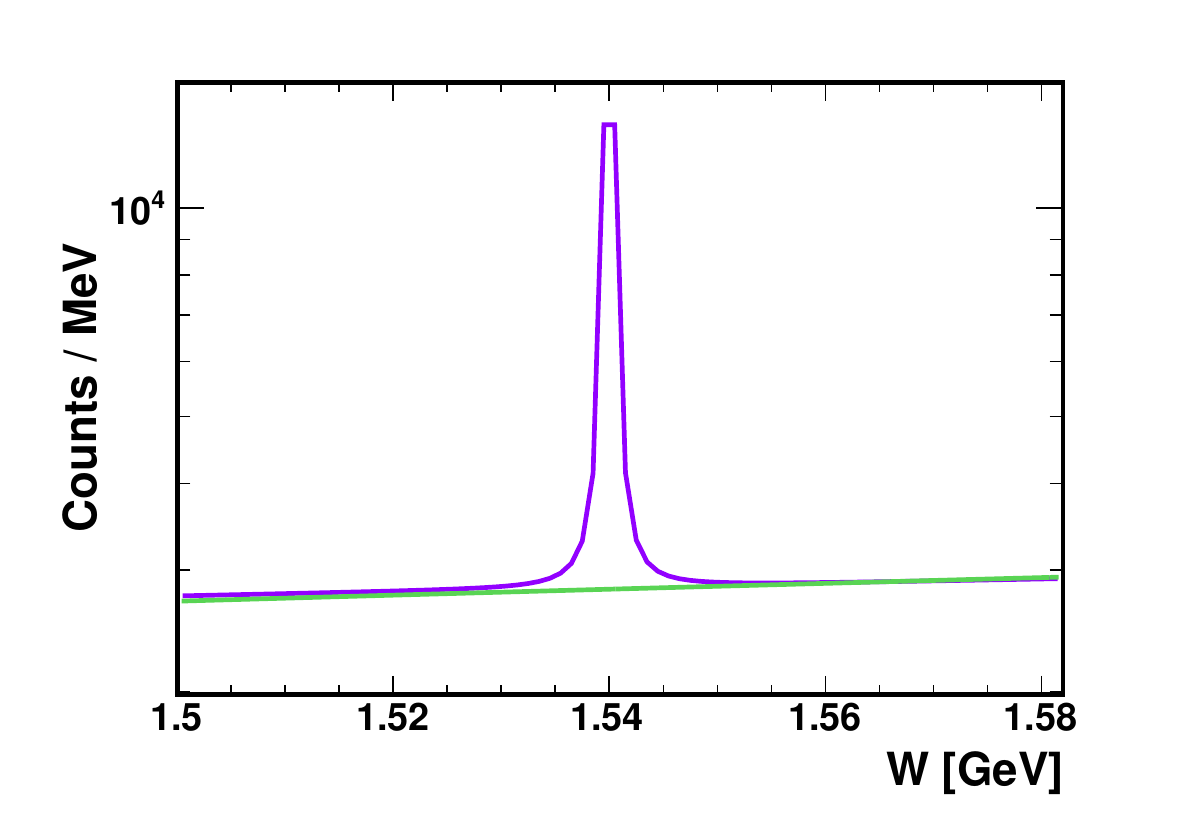} 
}

\centerline{\parbox{0.4\textwidth}{
\caption[] {\protect\small
Expected number of events in the reaction $K_Lp\to K^+n$ as a function of $W$. The background for $K_L p\to K^+ N$ (green solid curve) was simulated based on the model predictions~\cite{Iizawa:2023xsi}. The number of events in the peak for 100~days of running (purple solid curve) is estimated to be about 10,000~events (see text for details).  } 
\label{fig:res1} } }
\end{figure}

Reference~\cite{Iizawa:2023xsi} revisited the low-energy 
$K^+N$ elastic scatterings in the context of the in-medium quark condensate with strange quarks. There it described the KN amplitudes using chiral perturbation theory and fixed the low energy constants appearing in the amplitudes by existing KN scattering data. It allows to determine a non-resonant background cross-section estimation for the reaction $K_L p \to K^+ n$ 
(Fig.~\ref{fig:res1}). Then the total cross section for the kaon beam momentum in the laboratory frame $p_{K_L} = 0.440~\mathrm{GeV/c}$, which corresponds to $M_{\Theta^+} = 1.54~\mathrm{GeV}$,  is equal to $\sigma_{bkgd} = 3~\mathrm{mb}$.

The momentum resolution for $p_{K_L}$ at $440~\mathrm{MeV/c}$ is $\sigma_p/p = 6\times 10^{-3}$ (Fig.~\ref{fig:sigma1}), which is extremely important factor in a search for a resonance with a very narrow width.

Using Eq.~(1) from Ref.~\cite{DIANA:2013mhv} and Eq.~(3) from Ref.~\cite{Cahn:2003wq}, one can get a number of events in the peak as 
\begin{equation}
    N_{peak} = \frac{\Gamma(\Theta^+)~\pi\sigma_0~N_{bkgd}~B_iB_f}{2\sigma_{bkgd}~\Delta m_0} = 
    18,000~\mathrm{events} \>,
    \label{eq:eq2}
\end{equation}
where $N_{bkgd}$ corresponds to statistics during 100~days of the KLF running period ($N_{bkgd} = 5\times 10^3~\mathrm{events}$), 
$\Delta m_0$
is the $\Theta^+$ mass resolution corresponding to $\sigma_p/p$ and $\Delta m_0 = 1~\mathrm{MeV}$, branching ratios $B_i$ and $B_f$
into the initial and final channels of $\Theta^+$ according to the Breit-Wigner form (see Eq.~(1) from Ref.~\cite{Cahn:2003wq}) and $B_i = B_f = 1/2$. The $\sigma_0$ is a geometrical factor calculated as in Eq.~(2) from Ref.~\cite{Cahn:2003wq}
\begin{equation}
    \sigma_0 = \frac{2J + 1}{(2s_{K_L} + 1)(2s_p + 1)}\frac{4\pi}{k^2} = 68~\mathrm{mb} \>,
    \label{eq:eq1}
\end{equation}
where $k$ is the center-of-mass momentum of the  neutral kaon  beam ($k = 0.268~\mathrm{GeV/c}$), $s_{K_L}$ ($s_{K_L} = 0$) and $s_p$ ($s_p = 1/2$) are incident spins, and $J$ ($J = 1/2$) is the spin of the $\Theta^+$ $P_{01}$ resonance.

Finally, with all these calculations, we arrived to the number of  events of the $\Theta^+$ resonance in a $1~\mathrm{MeV}$ bin of the square root of invariant energy $W = s^{1/2}$, which is equal to the invariant mass of the 2-body $K^+N$ system. Thus, in 100~days of running of KLF it is expected to observe 18,000~events with the acceptance and efficiency correction it ends up to 10,000 events of the $\Theta^+$ formation  or impressive amount of 100 events per day. The corresponding plot is presented in Fig.~\ref{fig:res1}. 
%\textcolor{red}{It shows the invariant mass for $K^+n$ for
%exclusive two-body reaction $K_L p\to K^+n$. Monte Carlo shows that we identify the reaction by $K^+$s and undetected neutron will give us a distribution at about $50~\mathrm{MeV}$. That is good enough do not have a contribution for $K^+$ from different sources.}
It has to be mentioned that the statistics at KLF will exceed that obtained by DIANA 
experiment~\cite{DIANA:2013mhv} by $\sim$50~times.
%--------------------------------------------
%------------------------------------------------------------

%------------------------------------------------------------
\vspace{-0.5cm}
\section{Discussion and Outlook}
\label{Sec:Sum}
In summary, according to our estimation, about 10,000 events of the exotic $\Theta^+$ will be observed in a 100~days of running KLF. It is worth to mention that here we will measure not the invariant mass of $K^+n$ system, but rather the $W$ of the initial state for this reaction  benefiting from the extraordinary momentum resolution below $1~\mathrm{MeV}$ of the incoming neutral kaon momenta  in the region of interest.

Another proposal is developed, but not yet approved, to search for the $\Theta^+$ in the $K^+d \to K^0pp$ reaction at $p_{K^+} = 0.5~\mathrm{GeV/c}$ at J-PARC~\cite{Ahn:2023hiu}. The large acceptance Hyperon Spectrometer, which consists mainly of a time projection chamber and a $1~\mathrm{T}$ superconducting magnet, will exclusively measure the decay products of $\theta^+\to K^0p$ and $K^0\to \pi^+\pi^-$ decay. Although, a very interesting proposal, the final state is still a 3-body and one has to do model dependent calculations of the final-state interaction (FSI)~\cite{Migdal:1955ab, Watson:1952ji}, accurately take into account Fermi motion effects and avoid reflections from the combinatorial background. 

By comparing different reactions to measure directly the formation of the $\Theta^+$ it becomes clear that the unique way to make a formation of the $\Theta^+$ without any other associated particles is in reactions $K_Lp\to K^+n$ or $K_Lp \to K_Sp$. However, in the latter case strangeness is not fixed due to the presence of both $K^0$ and $\bar K^0$ in the wave function of  neutral kaons.

%------------------------------------------------------------------
%\vspace{1cm}
\section*{Acknowledgments}

%We thank Anatoly Dolgolenko, Eugene Chudakov, and Eulogio Oset for valuable comments and discussions.
The work of MJA was supported in part by the U.~S.~Department of Energy, Office of Science, Office of Nuclear Physics, under Award No.~DE--FG02-96ER40960. and the work of IIS in part under Award No.~DE--SC0016583. The work of DJ was supported in part by Grants-in-Aid for Scientific Research from JSPS (JP21K03530, JP22H04917, and JP23K03427).

%---------------------- REFERENCES ----------------------
%\input{references.tex}

%-----------------------------------------------
\end{document}